\documentclass[prd,superscripaddress,twocolumn,letterpaper,%
nofootinbid,preprintnumbers,amsmath,amssymb]{revtex4}

\usepackage{bm,curves}

\newcommand{\be}{\begin{equation}}
\newcommand{\ee}{\end{equation}}
\newcommand{\bear}{\begin{eqnarray}}
\newcommand{\ear}{\end{eqnarray}}
\newcommand{\la}{\label}

\begin{document}
\preprint{JLAB-THY-04-9}
\preprint{IFT-P.016/2004}
\title{Quark Distributions in a Medium}
\author{F. M. Steffens$^{a,b}$, A. W. Thomas$^c$, and K. Tsushima$^{a,b}$}
\affiliation{$^{a}$Mackenzie University - FCBEE, Rua da Consola\c{c}\~ao 930, 
01302-907, S\~ao Paulo - SP, Brazil}
\affiliation{$^{b}$Instituto de F\'{\i}sica Te\'orica - UNESP \linebreak
Rua Pamplona 145, 01405-900, S\~ao Paulo SP, Brazil}
\affiliation{$^c$Jefferson Lab, 12000 Jefferson Avenue, Newport News, VA 23606, U.S.A.}

\begin{abstract}
We derive the formal expressions needed to discuss the change of the
twist-two parton distribution functions when a hadron is placed in a
medium with relativistic scalar and vector mean fields.
\end{abstract}
\maketitle

\section{Introduction}

Presently, the quark distributions of the free proton are quite well 
known over a wide range of $x$ and $Q^2$.
However, the theoretical understanding of these distributions is  
somewhat limited. Although we 
know how to extract the quark distributions from the measured
structure functions within a NLO QCD analysis, we do not have sufficient
control of non-perturbative QCD to calculate these 
distributions from first principles. Nevertheless, we do know how
to formulate the problem in terms of the proton matrix elements
of certain local operators (see, for instance, Ref.~\cite{jaffe83}), which
correspond to the moments of the measured parton distributions. There
has been considerable progress in calculating at least the first few
moments using lattice QCD \cite{lattice} -- albeit at relatively large
quark masses~\cite{Detmold:2001jb}.

When it comes to nuclear structure functions one needs to evaluate the
matrix elements of these same operators in the nuclear ground state --
a priori a much more difficult problem. On the other hand, one knows
that nuclear structure functions (divided by the number of
constituent nucleons) lie within 10-20\% of the free nucleon structure
function (except near the kinematic boundary for the free nucleon) 
\cite{Reviews}. It therefore seems reasonable to tackle the problem
by computing the corrections to the structure function of a bound
nucleon and then allowing for Fermi motion. In particular, one could
consider as a starting point the case of infinite quark matter. Even
this presents serious theoretical challenges, because one knows from
numerous studies that one encounters large scalar and vector mean-fields
in this problem \cite{Relativity}, 
and there has been no discussion of the formal aspects of 
parton distributions in such an environment (analogous to the discussion
of Jaffe \cite{jaffe83} in the free case).

We therefore begin with a formal development of the parton model for the
case of a ``proton'' embedded in constant scalar and vector fields. Since
we are interested in modelling QCD, asymptotic freedom will be imposed
by hand, in that between the two hard collisions which define the
forward Compton amplitude (for the leading-twist parton distributions) 
the quark struck by the photon will be treated as free.
The result of this formal investigation is of general interest as the
formal properties of the parton distributions, in terms of support and
reflectivity or crossing symmetry ($x \rightarrow -x$), are guaranteed. 
Even so, it is possible to make suitable definitions of both
the valence and sea-quark distributions. In section II we derive the equations
for these quark distributions in a bound proton. In section III we verify that
they are normalized. Section IV is devoted to the investigation 
of the effects that the nuclear
medium exerts on the distributions, while in section V we apply our results
to the case of infinite, isospin symmetric, quark matter. Finally, in 
section VI we suggest directions for future work.

\section{Quark and Antiquark Distributions}

Our aim is to write down the formal expression for the 
quark and antiquark distributions 
of a bound proton. The in-medium proton momentum is 
denoted by $P^* = (P^{* 0}, \vec P^*)$, where 
the star superindex means that nuclear interactions 
have produced an effective mass and  energy for
the proton. As in deep inelastic scattering 
there are two independent variables to build the
hadronic tensor. We will use $P^*$ and $q$, 
the photon momentum probing the nuclei, 
as those variables. Hence, the hadronic tensor for 
the bound proton is written as:

\bear
&&W_{\mu\nu}(P^*, q) = (-g_{\mu\nu} + \frac{q_\mu q_\nu}{q^2})F_1^{BP}(x^*, q^2) \nonumber \\*
&+& (P^*_\mu - \frac{P^*\cdot q}{q^2}q_\mu ) (P^*_\nu - \frac{P^*\cdot q}{q^2}q_\nu)
\frac{F_2^{BP}(x^*, q^2)}{P^*\cdot q} \nonumber \\*
\la{e1}
\ear
where $F_1^{BP}(x^*, q^2)$ and $F_2^{BP}(x^*, q^2)$ are the structure functions
for the bound proton, and $x^* = -q^2/2P^*\cdot q$ is the fraction of the bound 
proton momentum carried by the quarks. In the parton model, the structure
functions are written as:

\bear
&&2x^* F^{BP}_1 (x^*, Q^2) = F^{BP}_2 (x^*, Q^2) \nonumber \\* 
&=& \sum_{i=u,d,s,...}e_i^2 (q_i^{BP}(x^*) + \overline q_i^{BP}(x^*)) + 
O(\alpha_s (Q^2)),
\la{e2}
\ear
where $Q^2 = -q^2$, $q_i^{BP}(x^*) (\overline q_i^{BP}(x^*))$ is the quark (antiquark) 
distribution in the bound proton and $O(\alpha_s (Q^2))$ are the QCD corrections to the parton model. 

In its simplest form, the quark distributions in the parton model are 
calculated from the handbag 
diagram. In the light-cone gauge, $A^+=0$, the distributions can be written 
as \cite{jaffe83}:
\bear
q^{BP} (x^*) &=& \frac{P^{*+}}{4\pi}\int dz^- e^{-ix^*P^{*+} z^-} \nonumber \\* 
& &<P^*|\psi^\dagger_+ (z^-) \psi_+ (0)|P^*>_c|_{z^+ = z^\perp = 0},
\la{e4}
\ear
where $P^{*+} = (P^{* 0} + P^{* 3})/\sqrt{2}$ is the plus component of the 
bound proton momentum  and $|P^*>$ is the state vector of the bound proton.
If the quark field operators in Eq.~(\ref{e4}) are expanded in terms 
of free plane waves, $q^{BP} (x)$ can be rewritten as:
\bear
q^{BP}(x^*) &=& \frac{P^{*+}}{2\pi}\int \frac{d^3 k}{(2\pi)^3}   
U_{\alpha \beta} (\vec k) \int dz^- e^{-ix^* P^{*+}z^- - i\vec k \cdot \vec z} 
\nonumber \\  
& & <P^*|b^{\alpha \dagger}_{\vec k} (t) b_{\vec k}^\beta (0)
|P^*>_c|_{z^+ = z^\perp = 0} \nonumber \\
&+&\frac{P^{*+}}{2\pi}\int \frac{d^3 k}{(2\pi)^3}
V_{\alpha \beta} (\vec k) \int dz^- e^{-ix^* P^{*+}z^- + i\vec k \cdot \vec z} 
\nonumber \\
& & <P^*|d_{\vec k}^\alpha (t) d^{\beta \dagger}_{\vec k}(0) 
|P^*>_c|_{z^+ = z^\perp = 0}, \nonumber \\
\la{e5}
\ear
where
$U_{\alpha \beta} (\vec k) = u^\dagger_{(\alpha)} 
(\vec k)[1 + \gamma^0 \gamma^3]
u_{(\beta)} (\vec k)$, $V_{\alpha \beta}(\vec k) =
v^\dagger_{(\alpha)} (\vec k)[1 + \gamma^0 \gamma^3]v_{(\beta)} (\vec k)$,
with $\alpha, \beta$ the polarization indices and the Einstein 
convention for repeated indices is understood. 
The time dependence 
of the creation and annihilation operators is calculated from:
\begin{equation}
b^{\alpha}_{\vec k} (t) = e^{i\hat H t}b^{\alpha}_{\vec k}e^{-i\hat H t},
\label{e6}
\end{equation}
with $\hat H$ the QCD Hamiltonian operator. We assume that the state 
vector $|P^*>$ is an eigenstate of $\hat H$ with eigenvalue $P^{* 0}$. 
Hence, the insertion of a complete set of intermediate states in Eq.~(\ref{e5}) 
implies that:
\bear
&& \hspace{-0.5cm} q^{BP}(x^*) = \nonumber \\*
&& \hspace{-0.5cm} \frac{P^{*+}}{2\pi}\int \frac{d^3 k}{(2\pi)^3}U_{\alpha \beta} (\vec k) 
\int dz^- e^{-ix^* P^{*+}z^- + i(P^{* 0} - P_n^{* 0})t - i\vec k \cdot\vec z} 
\nonumber \\*
&& \hspace{-0.5cm} \sum_n <P^*|b^{\alpha \dagger}_{\vec k}| n><n| b_{\vec k}^\beta|P^*>_c|_{z^+ = z^\perp = 0} \nonumber \\
&& \hspace{-0.5cm} + \frac{P^{*+}}{2\pi}\int \frac{d^3 k}{(2\pi)^3}V_{\alpha \beta}(\vec k)
\int dz^- e^{-ix^* P^{*+}z^- - i(P_n^{* 0} - P^{* 0})t +  i\vec k\cdot \vec z} 
\nonumber \\*
& & \hspace{-0.5cm} \sum_n <P^*|d_{\vec k}^\alpha |n><n| d^{\beta \dagger}_{\vec k}|P^*>_c|_{z^+ = z^\perp = 0},
\la{e7}
\ear
where $P_n^{* 0}$ is the eigenvalue of the intermediate 
state $|n>$ after the action of the
Hamiltonian operator. The integrals in $z^-$ in
Eq.~(\ref{e7}) can be done:
\bear
& &\int dz^- e^{-ix^* P^{*+}z^- + i(P^{* 0} - P_n^{* 0})t - i\vec k \cdot\vec z}|_{z^+ =  z^{\perp} = 0} \nonumber \\*
&=& \frac{2\pi}{P^{*+}} \delta(x^* - k_q^{* +}/P^{*+}),
\la{e8}
\ear
\bear
& &\int dz^- e^{-ix^* P^{*+}z^- - i(P_n^{* 0} - P^{* 0})t + i\vec k \cdot\vec z}|_{z^+ = z^{\perp} = 0} \nonumber \\*
&=& \frac{2\pi}{P^{*+}} \delta(x^* + k_{\overline q}^{* +}/P^{*+}),
\la{e9}
\ear
where $k_q^{* +} = (P^{* 0} - P_n^{* 0} + k_q^3)/\sqrt{2}$ is 
the plus component for quarks  
and $k_{\overline q}^{* +} = ( P_n^{* 0} - P^{* 0} + 
k_{\overline q}^3)/\sqrt{2}$ for antiquarks. 
We then have:
\bear
q^{BP}(x^*) &=&\int \frac{d^3 k}{(2\pi)^3}   
U_{\alpha \beta} (\vec k) \delta(x^* - k_q^{* +}/P^{*+}) \nonumber \\*
& & \sum_n <P^*|b^{\alpha \dagger}_{\vec k}|n><n| b_{\vec k}^\beta|P^*>_c \nonumber \\*
&+& \int \frac{d^3 k}{(2\pi)^3}   
V_{\alpha \beta} (\vec k) \delta(x^* + k_{\overline q}^{* +}/P^{*+}) \nonumber \\*
& & \sum_n <P^*|d^{\alpha}_{\vec k}|n><n| d_{\vec k}^{\beta \dagger}|P^*>_c.
\la{e10}
\ear
We note that the second term contributes to the quark distribution
because an antiquark with a negative momentum fraction can be
interpreted as a quark with positive momentum, $k_q^{* +} = -
k_{\overline q}^{* +}$.

If we assume that quarks in the nuclear environment feel scalar and
vector mean fields then we can write the vector potential as a shift 
in the quark and antiquark energies:
\begin{eqnarray}
k_q^{* +} &=& k_q^{+} + V^+ \nonumber \\*
k_{\overline q}^{* +} &=& k_{\overline q}^{+} - V^+,
\end{eqnarray}
where $V^+$ is the plus component of the vector potential, 
and $k_q^+$ ($k_{\overline q}^+$) the 
plus component of quark (antiquark) momentum with 
masses modified by the scalar potential. Hence, from  
delta functions in Eq. (\ref{e10}) we have:
\begin{equation}
x^* P^{*+} = x P^+ + V^+ \, ,
\la{e101}
\end{equation}
where $x = k_q^+/P^+$ is the fraction of momentum carried 
by a quark inside a proton immersed in
a medium with scalar but with no vector mean field. 
With the help of Eq.~(\ref{e101}) we can
isolate the effect of the vector potential on the quark distributions:
\begin{equation}
q^{BP}(x^*) = (P^{*+}/P^+) q^{BP}(x),
\label{e102}
\end{equation}
with the bound quark distribution without the vector potential defined as:
\begin{eqnarray}
q^{BP} (x) &=& \frac{P^{+}}{4\pi}\int dz^- e^{-ixP^{+} z^-} \nonumber \\* 
& &<P^*|\psi^\dagger_+ (z^-) \psi_+ (0)|P^*>_c|_{z^+ = z^\perp = 0}.
\label{e103}
\end{eqnarray}
As the effect of the vector mean field only changes 
the phase of the eigenstates, we 
conserved the notation $|P^*>$ for the bound proton 
with no vector potential when defining $q^{BP}(x)$. 
{}Finally, because the scalar potential can be 
absorbed in the proton mass, it follows 
that the whole analysis of Jaffe \cite{jaffe83} for 
the support of the Bjorken $x_{Bj}$ can be 
translated to the present case, 
meaning that the support for $x$ is $0 \leq x \leq 1$.
Similar relations to Eqs. (\ref{e101}) and (\ref{e102}) were also obtained by 
Mineo et al. \cite{Mineo:2003vc}, where the vector potential was 
treated as a gauge transformation.

On the other hand, the antiquark distribution in the parton model is given 
by the following expression \cite{jaffe83}:
\bear
\overline q^{BP} (x^*) &=& -\frac{P^{*+}}{4\pi}\int dz^- e^{-ix^* P^{*+} z^-} \nonumber \\*
& & <P^*|\psi^\dagger_+ (0) \psi_+ (z^-)|P^*>_c|_{z^+ = z^\perp = 0}.
\nonumber \\
\la{e11}
\ear
Inserting a complete set of intermediate states and 
performing the integrals over  
the $z^-$ variable, as in Eqs.~(\ref{e8}) and (\ref{e9}), we find the 
following expression for the antiquark distributions:
\bear
\overline q^{BP}(x^*) &=& \int \frac{d^3 k}{(2\pi)^3}   
U_{\alpha \beta} (\vec k) \delta(x^* + k_q^{* +}/P^{*+}) \nonumber \\*
& & \sum_n <P^*|b^{\alpha}_{\vec k}|n><n| b_{\vec k}^{\beta \dagger}|P^*>_c \nonumber \\*
&+&\int \frac{d^3 k}{(2\pi)^3}   
V_{\alpha \beta} (\vec k) \delta(x^* - k_{\overline q}^{* +}/P^{*+}) \nonumber \\*
& & \sum_n <P^*|d^{\beta \dagger}_{\vec k}|n><n| d_{\vec k}^\alpha|P^*>_c.
\la{e12}
\ear

The expressions for the quark and antiquark distributions, 
as given in Eqs.~(\ref{e10}) and (\ref{e12}), have (as they should) the crossing symmetry: 
$q^{BP}(x^*) = - \overline q^{BP}(-x^*)$. 

\section{Normalization}
The quark distributions of the bound proton must have the 
correct normalization. 
That is, the integral of $q^{BP} (x^*) - \overline q^{BP}(x^*)$ over the 
allowed $x^*$ range must give the number of valence quarks of the bound proton. 
To this end, we will use Eq.~(\ref{e101}), 
together with the allowed support for 
$x$. It follows that the maximum value for $x^*$ is $1 - 2 V^{+}$,
while the minimum value of $x^*$ is $V^{+}$:
\bear
&& \hspace{-0.5cm} \int_{V^+}^{1 - 2 V^+}[q^{BP} (x^*) - \overline q^{BP} (x^*)] dx^* = \nonumber \\*
&& \hspace{-0.5cm} \frac{P^{*+}}{2\pi} \int_{-\infty}^{\infty} dx^* dz^- e^{-ix^* P^{*+} z^-}
<P^*|\psi^\dagger_+ (z^-) \psi_+ (0)|P^*>_c \nonumber \\
&& \hspace{-0.5cm} = <P^*|\psi^\dagger_+ (0) \psi_+ (0)|P^*>_c \, ,  
\la{e14}
\ear
which gives the quark number in the state $|P^*>$:
\begin{equation}
\int_{V^+}^{1 - 2 V^+} [q^{BP} (x^*) - \overline q^{BP}(x^*)]dx^* = N^{BP}_{q} - N^{BP}_{\overline q}.
\la{e15}
\end{equation}
\section{The in-Medium Effects on the Distributions}

The quark and antiquark operators appearing in the distributions (\ref{e10}) and
(\ref{e12}) are those of free quantum fields. The quarks which build the proton
state are, however, not in free space: 
they are confined in a proton state, while 
the proton state is itself immersed in a nuclear medium. 
Our state vector has to be built
from these bound quark operators. 
We shall denote by $b_{\vec p}^*$ ($d_{\vec p}^*$) 
the annihilation operator of a quark (antiquark) in the bound proton. 
Let the bound proton state be written as:
\begin{equation}
|P^*> = F[q_{\vec p_1}^{* \dagger},q_{\vec p_2}^{* \dagger},q_{\vec p_3}^{* \dagger}]|0^*>,
\la{e16}
\end{equation}
where 
$F[q_{\vec p_1}^{* \dagger},q_{\vec p_2}^{* \dagger},q_{\vec p_3}^{* \dagger}]$ 
is a functional of the bound quark operators, $|0^*>$ is the effective vacuum 
(as seen by quarks in a nuclear medium) 
where the quarks bound in the proton live, 
and $\vec p_1, \vec p_2, \vec p_3$ are the individual momenta 
of the three valence quarks. 
Although only the valence quarks appear explicitly in the functional, 
it is understood 
that it may be populated by quark - antiquark pairs and gluons: the notation 
only shows the net number of quarks and antiquarks.
  
The effective vacuum is defined by:
\begin{equation}
b_{\vec p}^*|0^*> = 0 \;\;,\;\; d_{\vec p}^*|0^*> = 0.
\la{e17}
\end{equation}
With this definition, we can calculate the action of the free 
quark and antiquark operators
on the bound proton state. 
To do this, note that the quark field operator in Eq.~(\ref{e4})
could have been expanded in any basis \cite{kazuo01}. 
Suppose that we know the solution of the Dirac 
equation for the interacting theory describing the bound proton 
state, with solutions $u_{BP}(\vec p,\vec x)$ ($v_{BP}(\vec p,\vec x)$) 
for the positive (negative) energy part. The
field operator expanded in this basis is written as:
\begin{equation}
\psi(x) = \int \frac{d^3p}{(2\pi)^3} \sum_\alpha [b^*_{\vec p ,\alpha}(t) u_{BP}^{(\alpha)}(\vec p,\vec x)
+ d^{*\dagger}_{\vec p, \alpha}(t) v_{BP}^{(\alpha)}(\vec p,\vec x)],
\la{e18}
\end{equation}
with $u^\dagger u = 1$ \cite{Relativity}.
%E^*/m^* = \sqrt{(m^*)^2 + \vec p^2}/m^*$, and $m^*$ the 
%mass of the quark in the bound proton.
%
If we compare it with the expansion in terms of the free fields, we get:
\bear
\hspace{-0.5cm}b_{\vec k,\alpha}(t) &=& \int \frac{d^3p}{(2\pi)^3} \nonumber \\*
& &\sum_\beta [b^*_{\vec p,\beta}(t) A^{\alpha\beta}(\vec k,\vec p)
+ d^{*\dagger}_{\vec p,\beta}(t) B^{\alpha\beta} (\vec k,\vec p)],
\la{e90}
\ear
where:
\begin{equation}
A^{\alpha\beta}(\vec k,\vec p) = \int d^3x \overline u^{(\alpha)}(\vec k)\gamma_0 u_{BP}^{(\beta)}(\vec p,\vec x) 
e^{-i\vec k\cdot \vec x},
\la{e20}
\end{equation}
\begin{equation}
B^{\alpha\beta} (\vec k,\vec p) = \int d^3x \overline u^{(\alpha)}(\vec k)\gamma_0 v_{BP}^{(\beta)}(\vec p,\vec x) 
e^{-i\vec k\cdot \vec x}.
\la{e21}
\end{equation}
The calculation of the anticomutator between the free and bound operators gives:
\begin{equation}
\{b_{\vec k}^\alpha (t), b_{\vec p}^{* \beta \dagger} (t^\prime)\}_{t=t^\prime} = A^{\alpha\beta}(\vec k , \vec p),
\la{e22}
\end{equation}
and similar for the other operators. 
Thus the action of the free quark annihilation operator 
in the proton state results in:
\bear
b_{\vec k}^\beta|P^*> &=& A^{\beta\gamma}(\vec k , \vec p_i)G[b_{\vec p_j}^{* \dagger},
b_{\vec p_l}^{* \dagger}]^{\gamma}|0^*> \nonumber \\*
&-& F[b_{\vec p_1}^{* \dagger},b_{\vec p_2}^{* \dagger},b_{\vec p_3}^{* \dagger}] b_{\vec k}^\alpha
|0^*>,
\la{e23}
\ear
where $G[b_{\vec p_j}^{* \dagger},b_{\vec p_l}^{* \dagger}]^{\gamma}$ is some function of 
the effective quark operators after 
$b_{\vec k}^\alpha$ acts on $F$, and the 
index $\gamma$ indicates that the resulting two 
quark states are in coloured states. 
Integration over $\vec p_i$, $\vec p_j$ and $\vec p_l$ 
is understood. Thus, the action of the free operator on the effective
vacuum yields:
\begin{equation}
b_{\vec k}^\beta |0^*> = \int \frac{d^3p}{(2\pi)^3}\sum_\gamma B^{\beta\gamma}(\vec k,\vec p)
|\overline{q}_{\vec p,\gamma}^*>.
\la{e24}
\end{equation}
{}From Eqs.~(\ref{e23}) and (\ref{e24}) we see that 
there will be contributions from 
two terms in the quark distribution, Eq.~(\ref{e10});  
one when we have two quarks
in the intermediate state and the other when we have three 
quarks and one antiquark in 
the intermediate state. For the case $n=2$, we define the coloured state:
\begin{equation}
|n=2> = G[b_{\vec p_j}^{\dagger *},
b_{\vec p_k}^{\dagger *}]^\gamma |0^*>,
\la{e25}
\end{equation}
and for the case $n=4$:
\begin{equation}
|n=4> = F[b_{\vec p_1}^{\dagger *},b_{\vec p_2}^{\dagger *},b_{\vec p_3}^{\dagger *}]
|\overline{q}_{\vec p,\gamma}^*>.
\la{e26}
\end{equation}
In a completely analogous way to what was done 
in Eqs.~(\ref{e90}) - (\ref{e21}) for 
the free quark operator, 
we can express the free antiquark operator in terms of the 
bound quark and antiquark operators. 
We will call $C^{\alpha\beta} (\vec k,\vec p)$ the overlap 
analogous to the  $B^{\alpha\beta} (\vec k,\vec p)$ term, 
with the interchange between the 
$u$ and $v$ Dirac spinors. 
Similarly, $D^{\alpha\beta} (\vec k,\vec p)$ will be the analogue  
of the $A^{\alpha\beta} (\vec k,\vec p)$ overlap, with the $v$ Dirac spinor replacing the 
$u$ Dirac spinor everywhere. 

Using the states defined in Eqs.~(\ref{e25}) and (\ref{e26}), 
the quark distribution 
Eq.~(\ref{e10}) is finally written as:
\bear 
q^{BP}(x^*) &=& \int \frac{d^3 k}{(2\pi)^3} \frac{d^3 p}{(2\pi)^3}   
U_{\alpha \beta} (\vec k) \delta (x^* - k_q^{* +}/P^{*+}) \nonumber \\
& &\sum_{\gamma}\left( <n=2|n=2 > A^{\gamma \alpha \dagger}(\vec k, \vec p ´) A^{\gamma\beta}(\vec k ,\vec p) \right. \nonumber \\
&-& \left.<n=4|n=4 > B^{\gamma \alpha \dagger}(\vec k,\vec p') B^{\gamma\beta}(\vec k,\vec p)\right) \nonumber \\* 
&+& 
\int \frac{d^3 k}{(2\pi)^3} \frac{d^3 p}{(2\pi)^3}  
V_{\alpha \beta} (\vec k) \delta (x^* + k_{\overline q}^{* +}/P^{*+}) \nonumber \\
& &\sum_{\gamma}\left( <n=2|n=2 > D^{\gamma \alpha \dagger}(\vec k, \vec p ´) D^{\gamma\beta}(\vec k ,\vec p) \right. \nonumber \\
&-& \left.<n=4|n=4 > C^{\gamma \alpha \dagger}(\vec k,\vec p') C^{\gamma\beta}(\vec k,\vec p)\right)
\nonumber \\ 
\label{e27}
\ear
where integration over all the internal momenta of the 
quarks in the intermediate states
is understood.

\section{The Quark Matter Case}

A central question from the point of view of nuclear physics involves
the changes to the quark and antiquark distributions of a bound proton. 
Since one must develop a reliable model of both the free proton and the
binding of nucleons starting from the 
quark level~\cite{Saito:1992rm}, this problem is
rather complicated. We intend to report on our investigation of that
problem in future work. For the present, we have chosen to illustrate
the formal ideas developed here by applying them to a toy model, namely 
the quark distributions of isospin symmetric quark matter in which each
quark feels a 
scalar potential, $- V_s^q$, and a vector potential, $V_v^q$. 
This is the premise of the 
Quark Meson Coupling (QMC) model~\cite{Guichon:1987jp} which has been used 
successfully to calculate the 
properties of nuclear matter as well as finite
nuclei~\cite{Saito:1996yb,Saito:ki}. Most recently it has also been used 
to derive an effective nuclear force which is very close to the widely used 
Skyrme III force~\cite{Guichon:2004xg}.
(Except that in QMC the quarks are confined by the MIT bag, as well as
feeling the mean-field scalar and vector potentials
generated by the surrounding nucleons.) In the mean field approximation, 
the Dirac Equation for the quark in infinite quark matter is written as:
\begin{equation}
[i\gamma\cdot\partial - (m - V^q_s) - \gamma_0 V^q_v]\psi_{QM}^q(x)=0.
\la{e28}
\end{equation}

{}Following Eq.~(\ref{e18}), we write the field operator in terms of the 
solutions of Eq.~(\ref{e28}):
\bear
\psi^q_{QM} (x) &=& \int \frac{d^3 p}{(2\pi)^3} \sum_\alpha \nonumber \\*
& & \left[b^*_{\vec p ,\alpha} u^{(\alpha)}_{QM}(\vec p) 
 e^{-i p_q^{* 0}t + i \vec p\cdot\vec x} \right. \nonumber \\*
& &\;\;\; + \left. d^{* \dagger}_{\vec p ,\alpha} 
 v^{(\alpha)}_{QM}(\vec p) e^{i p_{\overline q}^{* 0}t - i \vec p\cdot\vec x}\right], 
\la{e29}
\ear
where $u_{QM}^{(\alpha)}(\vec p)$ and $v_{QM}^{(\alpha)}(\vec p)$ 
are the in-medium Dirac spinors with 
mass $m^* = m - V^q_s$, energy $E_q^*$, and 
where $p^{* 0}_q = E_q^* + V_v^q$ for quarks 
and $p^{* 0}_{\overline q} = E_q^* - V_v^q$ for antiquarks. 

We now calculate Eq.~(\ref{e20}) for the $A$ factor, 
using $k = (p_k^0, \vec k)$ for the 
free quark and $p=(p_q^{* 0},\vec p)$ for the quark in infinite quark matter:
\bear
& &A^{\alpha\beta}(\vec k,\vec p) = \nonumber \\*
&& \int d^3x \overline u^{(\alpha)}(\vec k)\gamma_0 u_{QM}^{(\beta)}(\vec p) 
e^{i(\vec p - \vec k)\cdot \vec x }.
\la{e32}
\ear
The integral over $\vec x$ can be done, giving a Dirac delta function, 
which implies that:
\begin{equation}
A^{\alpha\beta}(\vec k,\vec p) = (2\pi)^3 \delta(\vec p - \vec k) \overline u^{(\alpha)} (\vec k)\gamma_0 u_{QM}^{(\beta )}(\vec p).
\la{e33}
\end{equation}
Similarly, Eq. (\ref{e21}) for the $B$ factor becomes:
\begin{equation}
B^{\alpha\beta}(\vec k,\vec p) = (2\pi)^3 \delta(\vec p + \vec k)\overline u^{(\alpha)} (\vec k)\gamma_0 v_{QM}^{(\beta )}(\vec p).
\la{e331}
\end{equation}
If we go back to Eq.~(\ref{e90}) relating the free and 
bound quark operators, we find:
\bear
b_{\vec k}^{\alpha}(t) &=& \overline u^{(\alpha)} (\vec k)\gamma_0 u_{QM}^{(\beta )}(\vec k) b^*_{\vec k ,\beta}(t) \nonumber \\*
& & + \overline u^{(\alpha)} (\vec k)\gamma_0 v_{QM}^{(\beta )}(-\vec k) d_{-\vec k,\beta}^{* \dagger}(t).
\la{e34}
\ear

A similar calculation for the antiquark operator gives:

\bear
d_{\vec k}^{\alpha \dagger}(t) &=& \overline v^{(\alpha)} (\vec k)\gamma_0 u_{QM}^{(\beta )}(-\vec k) 
b^{* \beta}_{-\vec k}(t) \nonumber \\*
& & + \overline v^{(\alpha)} (\vec k)\gamma_0 v_{QM}^{(\beta )}(\vec k) d_{\vec k,}^{*\dagger \beta}(t).
\la{e341}
\ear
An explicit calculation shows that 
$D^{\alpha\beta}(\vec k, \vec p) = A^{\alpha\beta}(\vec k, \vec p)$ and
$C^{\alpha\beta}(\vec k, \vec p) = B^{\alpha\beta}(\vec k, \vec p)$, 
which implies that charge conjugation holds between the bound quark and antiquark
operators.

Although the quark and antiquark operators in Eqs. (\ref{e34}) and (\ref{e341}) are time
dependent, the expressions for the quark distributions involve products like $b^\dagger b$,
which are time independent:

\begin{equation}
b_{\vec k}^\dagger b_{\vec k} = b_{\vec k}^\dagger (t) b_{\vec k}(t).
\la{e342}
\end{equation}
The bound quark distribution, calculated from Eq.~(\ref{e10}), is then:

\bear
q^{BP}(x^*) &=& \int \frac{d^3 k}{(2\pi)^3}
U^{\alpha \beta}(\vec k)\delta(x^* - k_q^{* +}/P^{*+}) \nonumber \\*
& & \;\;\;\;\; \times  \frac{E_q + m}{2 E_q} \frac{E_q^* + m^*}{2 E_q^*} \nonumber \\*
& &\hspace{-1.5cm} \times\left( \left[ 1 + \frac{\vec k^2}{(E_q + m) (E_q^* + m^*)}\right]^2 
<P^*|b^{*\dagger}_{\vec k, \alpha}b^{*}_{\vec k, \beta}|P^*>\right. \nonumber \\*
& & \hspace{-1.1cm} + \left. \left[ \frac{1}{E_q + m} - \frac{1}{E_q^* + m^*}\right]^2 \right. \nonumber \\*
& & \hspace{-1cm} \times \left. (\chi_\gamma^\dagger \vec\sigma\cdot\vec k \chi_\alpha) 
(\chi_\beta^\dagger \vec\sigma\cdot\vec k \chi_\delta)
<P^*|d_{-\vec k}^{* \gamma}d_{-\vec k}^{* \delta \dagger}|P^*> \right) 
\nonumber 
\ear
\bear
&+& \int \frac{d^3 k}{(2\pi)^3}
V^{\alpha \beta}(\vec k)\delta(x^* + k_{\overline q}^{* +}/P^{*+}) \nonumber \\*
& & \;\;\;\;\; \times  \frac{E_q + m}{2 E_q} \frac{E_q^* + m^*}{2 E_q^*} 
\nonumber \\*
& & \hspace{-1.5cm} \times \left( \left[ 1 + \frac{\vec k^2}{(E_q + m) (E_q^* + m^*)}\right]^2 
<P^*|d^{*}_{\vec k, \alpha}d^{* \dagger}_{\vec k, \beta}|P^*> \right. \nonumber \\*  
& & \hspace{-1.1cm} + \left. \left[ \frac{1}{E_q + m} - \frac{1}{E_q^* + m^*}\right]^2 \right. \nonumber \\*
& & \hspace{-1cm} \times \left. (\chi_\gamma^\dagger \vec\sigma\cdot\vec k \chi_\alpha)
(\chi_\beta^\dagger \vec\sigma\cdot\vec k \chi_\delta)
<P^*|b_{-\vec k}^{* \gamma \dagger}b_{-\vec k}^{* \delta}|P^*> \right) \nonumber \\*
\label{e35}
\ear

\section{Concluding Remarks}
We have developed the formal framework for evaluating the change in the
structure function of a hadron when it is bound in a nuclear medium. In
particular, the formalism is able to deal with the separate 
changes in the quark and antiquark distributions, while preserving the
necessary sum rules.  
Since much of our information on the parton distribution functions of
the nucleon does in fact come from nuclear data this is an especially
important issue~\cite{Benesh:2003fk,Smith:ak,Thomas:2003ge}. 
In order to illustrate the formalism we considered the
parton distribution functions for quark matter with mean field scalar
and vector potentials. In the immediate future it will be important to
include the effect of confinement, as for example in 
the QMC model~\cite{Guichon:1987jp,qmc1} or
even the NJL model~\cite{Mineo:2003vc,Bentz:2002um} with proper time 
regularization~\cite{Ebert:1996vx,Hellstern:1997nv}.

\section*{Acknowledgements}
This work was supported by  CNPq (contract 308032/2003-0). 
It was also supported by the Australian Research Council and by 
DOE contract DE-AC05-84ER40150, under which SURA operates 
Jefferson Laboratory. KT is supported by FAPESP
contract 2003/06814-8.
\addcontentsline{toc}{chapter}{\protect\numberline{}{References}}

\end{document}